\documentclass[aps,prc,twocolumn,groupedaddress,showpacs,superscriptaddress,nofootinbib]{revtex4-1}
\usepackage{graphicx,amsmath}
\usepackage{amssymb}
\usepackage{hyperref}
\usepackage[version=3]{mhchem}
\newcommand{\qua}{$^4{\rm He}$}

\newcommand{\sep}{$^{7}{\rm Li}$}
\newcommand{\bery}{$^{7}$Be}

\newcommand{\ddn}{{D(d,n)$^3$He}}
\newcommand{\ddp}{{D(d,p)$^3$H}}
\newcommand{\dpg}{{D(p,$\gamma)^3$He}}
\newcommand{\hag}{{$^3$He($\alpha,\gamma)^7$Be}}

\newcommand{\sfac}{$S$--factor}
\newcommand{\zaa}{Astron.~Astrophys.}

\newcommand{\zapj}{Astrophys.~J.}
\newcommand{\zapjl}{Astrophys.~J.~Lett.}

\newcommand{\znp}{Nucl.~Phys.}

\newcommand{\zpr}{Phys.~Rev.}

\newcommand{\zprc}{ Phys.~Rev. C}
\newcommand{\zprl}{Phys.~Rev.~Lett.}

\newcommand{\zadndt}{At. Data Nucl. Data Tables}

\newcommand{\zjcap}{J. Cosmol. Astropart. Phys.}

\begin{document}


\title{Comment on ``Measurement of \emph{d} + \bery\ Cross Sections for Big-Bang Nucleosynthesis''}



\author{Alain Coc}
\affiliation{Centre de Sciences Nucl\'eaires et de Sciences de la Mati\`ere, CNRS/IN2P3,  B\^atiment, 104, F-91405 Orsay Campus, France}
\author{Barry Davids}
\affiliation{TRIUMF, 4004 Wesbrook Mall, Vancouver, BC, V6T 2A3, Canada}
\affiliation{Department of Physics, Simon Fraser University, 8888 University Drive, Burnaby, BC, V5A 1S6, Canada}

\begin{abstract}
In a recent Letter entitled ``Measurement of \emph{d} + \bery\ Cross Sections for Big-Bang Nucleosynthesis'', Rijal \emph{et al.} \cite{Rij19} 
describe the results of an experiment that put the \bery~+ \emph{d} reaction rate on firmer ground. 
The erratum \cite{Rij19a} claims that including this reaction rate  in big bang nucleosynthesis (BBN) brings a significant reduction of up to 8.1\% of lithium production relative to calculations that exclude this reaction.
In contrast, using several well tested BBN codes, we confirm earlier studies showing that the effect is less than 1\%, 
so that this measurement cannot alleviate the cosmological lithium problem.
\end{abstract}

\date{\today}


\maketitle


A Letter by Rijal \emph{et al.} \cite{Rij19} reports a previously unknown resonance within the Gamow window for big bang nucleosynthesis. 
The BBN calculations presented in Ref.\ \cite{Rij19} show a significant reduction 
of the primordial lithium abundance. (In BBN, 95\% of \sep\ comes from the subsequent decay of \bery.) However, this result cannot be 
reproduced by our well tested BBN codes \cite{CV17,Pit18,kawano92}, using evaluated reaction 
rates\footnote{In particular, the {\it PRIMAT} code \cite{Pit18}, freely available at {\tt http://www2.iap.fr/users/pitrou/primat.htm}}, 
first because  $i$) the new rate (hereafter the FSU rate) is close to the previous estimate, but more importantly because $ii$) calculations 
show that the rate is too low  to have any appreciable influence  \cite{omeg} on BBN Li production. The first point has recently been discussed in 
Ref.\ \cite{Gai19}; here we focus on the second, more crucial one.


BBN has entered the precision era for \qua\ and deuterium (see \emph{e.g.} Refs. \cite{Cyb16,Pit18}), but is plagued by the {lithium problem} \cite{Fie11,Cyb16,Pit18}: the predicted final \sep\ abundance is a factor of $\approx3$ higher 
than observations \cite{Sbo10}. Nuclear physics solutions to this problem have been extensively investigated \cite{Coc04,boyd10,broggini12,Coc12a,{cyburt12}}, showing, in particular, that if the \bery~+~\emph{d} reaction rate were much higher (a factor of $\sim$100) than the commonly used estimate of Caughlan and Fowler \cite{CF88} (hereafter the CF88 rate), then the problem would be alleviated.   

\begin{figure}[htb]
\begin{center}
 \includegraphics[width=.45\textwidth]{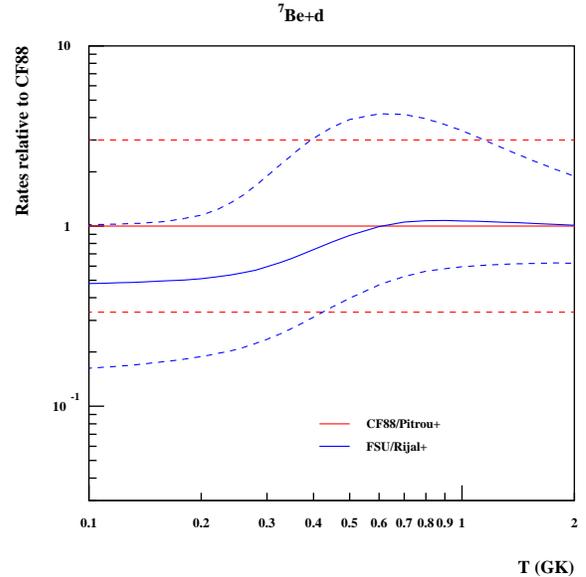} 
\caption{Ratio of the thermally averaged \bery+\emph{d} reaction rate of Ref.\ \cite{Rij19} (FSU) to the recommended rate of Ref.\ \cite{CF88} (CF88). The red line represents the CF88 rate with dashed upper and lower limits corresponding to a factor of 3 uncertainty \cite{Pit18}, while the blue curves correspond to the {\it Low}, {\it Mid}, and {\it High} rates from the Supplemental Material of Ref.\ \cite{Rij19}.
}
\label{f:bdarate}
\end{center}
\end{figure}

Indeed, the CF88 rate, used in previous BBN calculations (e.g. Ref. \cite{Pit18}), was based on partial measurements 
\cite{Kav60}, supplemented by ``educated guesses'' (a factor of 3 increase \cite{Par72,Gai19} and a factor of 3 
uncertainty \cite{Pit18}). A more recent experiment \cite{Ang05} was performed at BBN energies but was only sensitive to the  $^7$Be(d,p)$^8$Be(2$\alpha$) channel. Rijal \emph{et al.} \cite{Rij19} found a new resonance, in the previously unexplored $^7$Be(d,$\alpha$)$^5$Li(p$\alpha$) channel. The reported experimental \sfac\ is much higher than the previous ones \cite{Kav60,Ang05}, but, because of the overestimation resulting from the multiplicative factor introduced by Ref.\ \cite{Par72}, by pure chance the FSU rate is close to the CF88 rate \cite{Gai19}. This fact is only briefly mentioned  in Ref.\ \cite{Rij19}, in which emphasis is placed on a comparison with rates derived from the incomplete measurements of Refs.\ \cite{Kav60,Ang05} that have never before, to our knowledge, been used in BBN calculations. (One of us is a co-author of the Angulo \emph{et al.} paper \cite{Ang05} in which {\em no} reaction rate is given.) The factor of three uncertainty assumed in Ref.\ \cite{Pit18} for the CF88 rate is also very close to the quoted uncertainty of the FSU rate, as can be seen in Fig.~\ref{f:bdarate}.

The properties of a state in $^9$B that would be required for a significant resonant enhancement of the \emph{d} + \bery\ cross section at BBN energies were described in Ref.\ \cite{cyburt12}. Evidence for such a resonance was sought but not found in a  \bery\ + \emph{d} elastic scattering measurement \cite{OMa11}. The authors of Ref.\ \cite{Kir11} identified a state at 16.8~MeV in $^9$B, whose excitation energy and width were measured in Ref.\ \cite{Sch11}, as the only known candidate for such a resonance. However, accompanying BBN calculations ruled out any significant influence of the resonance on the predicted final $^7$Li abundance, even assuming the maximum possible resonance strength. It was pointed out in Ref.\ \cite{Kir11} that although the analysis was based on the assumption of a dominant $p$ decay branch of the 16.8~MeV state to the 16.626~MeV state in $^8$Be, similarly insufficient reductions in the final $^7$Li abundance were found when the 16.8~MeV state was assumed to decay dominantly by $\alpha$ emission. The resonance strength reported in Ref.\ \cite{Rij19} is much smaller than that considered in Ref.\ \cite{Kir11}. Hence it has long been known that a \bery\ + \emph{d} resonance with the properties reported in Ref.\ \cite{Rij19} cannot have the effect on $^7$Li abundance claimed by Rijal \emph{et al.}

\begin{figure}[htb]
\begin{center}
 \includegraphics[width=0.5\textwidth]{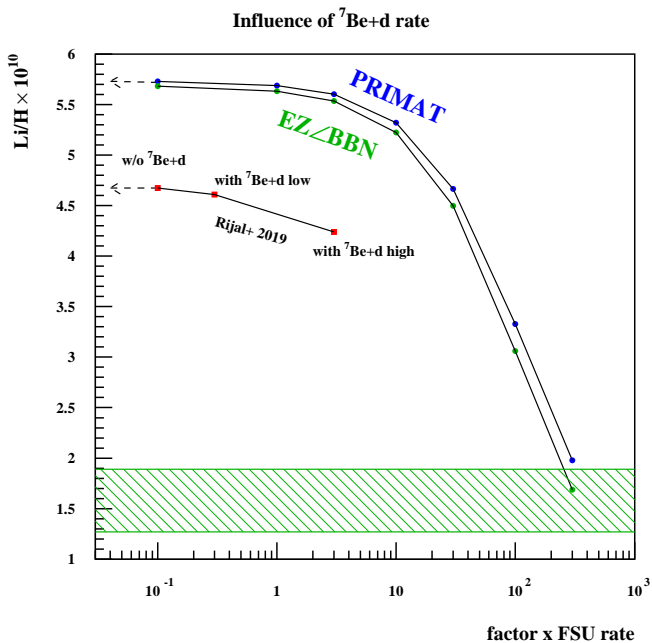} 
\caption{Dependence of final $^7$Li/H on the thermally averaged  \bery+d reaction rate in units of the FSU rate according to our calculations with the {\it PRIMAT} \cite{Pit18} and {\it EZ\_BBN} 
\cite{CV17} codes and according to Ref.\ \cite{Rij19}. (For convenience, the results without  the \bery~+~$d$ reaction are plotted at an arbitrary factor of 0.1. FSU results are plotted
at 0.3 and 3 factors for the lower and higher rate limits respectively, as an approximation based on ratios displayed on Fig.~\ref{f:bdarate}.) 
Calculations with the Kawano-Wagoner code \cite{kawano92} show a very similar trend. The hatched region represents the observations of Ref.\ \cite{Sbo10}.}
\label{f:fsu}
\end{center}
\end{figure}

\begin{figure}[htb]
\begin{center}
 \includegraphics[width=0.5\textwidth]{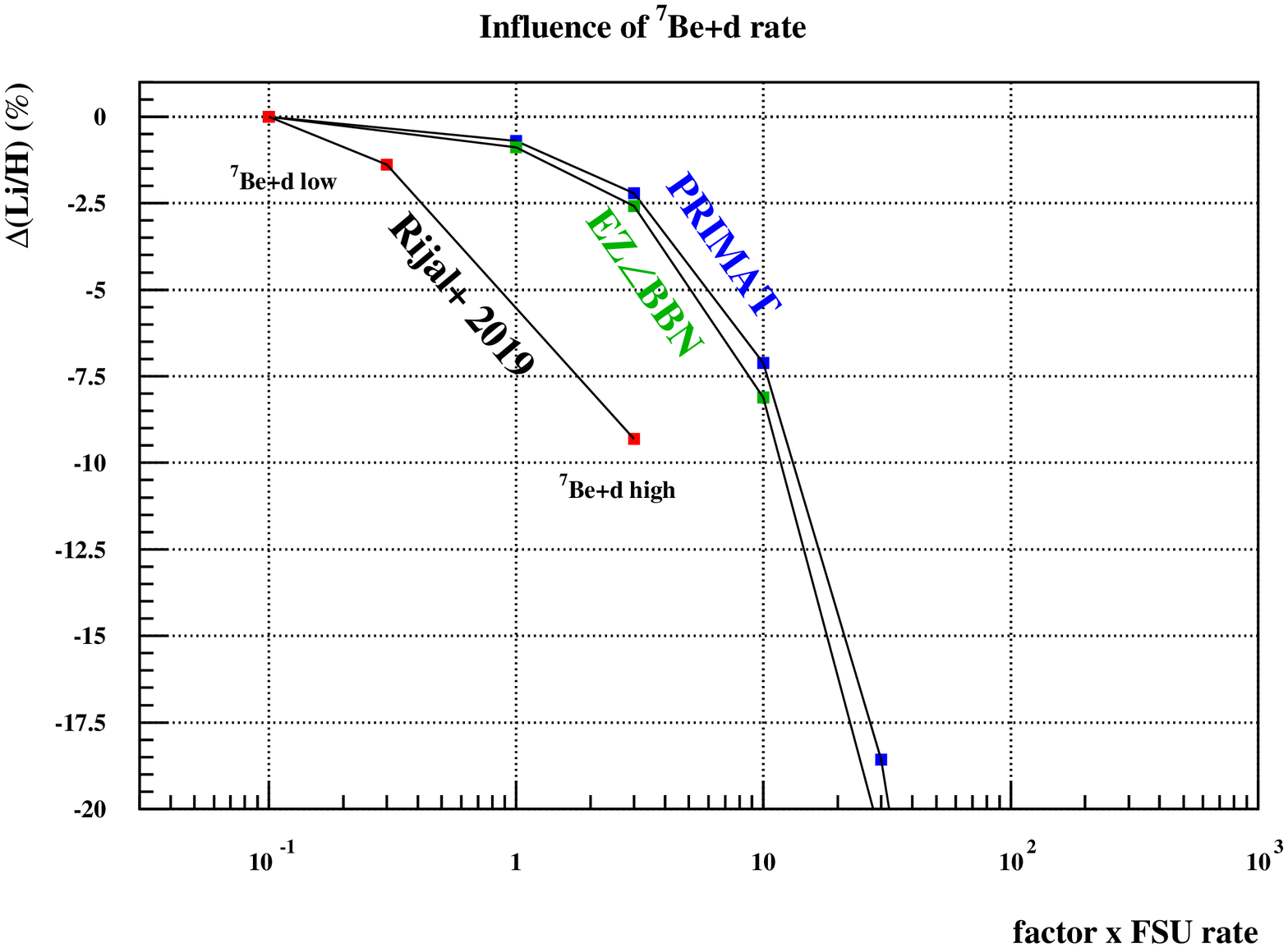} 
\caption{Same as Fig.~\ref{f:fsu}, but showing the percent variation of $^7$Li/H instead of the absolute values.}
\label{f:fsu2}
\end{center}
\end{figure}

Thanks to the Rijal \emph{et al.} experiment we are now confident in the reaction rate to be used, but as the FSU rate is not significantly different from the previously adopted CF88 rate, no appreciable reduction in the $^7$Li/H
prediction is expected. Indeed, when switching this reaction on and off in our BBN codes using the FSU rate we obtain a decrease in the final $^7$Li/H of 0.7\%, in conflict with the 1.4-8.1\% decrease reported in Ref.\ \cite{Rij19a}. It is worth noting that the Erratum underestimates the fractional reduction in primordial Li according to its own calculations. The first paragraph says the authors compute Li/H $\times10^{10} = 4.66 - 4.69$ for the case where the $^7$Be$ + d$ rate is neglected and Li/H $\times10^{10} = 4.24 - 4.61$ for the range of reaction rates they determine. These numbers imply an absurdly small uncertainty in the primordial Li abundance prediction when the $^7$Be$ + d$ rate is neglected. They also imply that by the authors' own calculations the reduction in Li/H for the upper limit of the Rijal \emph{et al.} reaction rate is 9.3\%, not the 8.1\% given in the Erratum \cite{Rij19a}.

When we compare the effect of using the FSU rate instead of the CF88 rate, the difference in the final $^7$Li/H value is much smaller than 0.7\%. This  confirms, as has long been known \cite{Coc04,Coc12a}, that an increase of the $^7$Be + \emph{d} rate by factor of $>$30 with respect to the CF88 rate is required to have a significant influence on the final $^7$Li abundance; Fig.~\ref{f:fsu} and Fig.~\ref{f:fsu2} show the effect of varying the $^7$Be + \emph{d} rate on the predicted $^7$Li abundance. We obtained very similar results with three different codes: the new {\it Mathematica} code {\it PRIMAT} \cite{Pit18}, a {\it Fortran 77} code, {\it EZ\_BBN}, used and improved extensively over the years \cite{CV17}, and a modified version of the Kawano-Wagoner code \cite{kawano92}. Moreover, our results are completely consistent with those of Ref.\ \cite{fields19}.
The difference in the absolute values of Li/H, seen in  Fig.~\ref{f:fsu}, when the  $^7$Be~+~\emph{d} reaction is switched off, can easily be explained by the use of
different reaction rates.  For instance, the difference between the Pitrou \emph{et al.}  \cite{Pit18} and Cyburt \emph{et al.}  \cite{Cyb16} BBN predictions vanishes \cite{private} if the same rates are used
for the  \dpg,  \ddn, \ddp\  and   \hag\ reactions.     
Since very few details are given on the BBN code used in Ref.\ \cite{Rij19}, we cannot explain why its results are so different from those of our well-tested codes.

In conclusion, we confirm that the inclusion of the  $^7$Be~+~\emph{d} reaction in BBN calculations does not affect the final $^7$Li/H prediction by more than $\approx$1\%,
hence not significantly enough to reduce the cosmological lithium problem.

\end{document}